\documentclass[12pt]{article}
\usepackage{amsmath,amssymb,exscale}
\usepackage{psfrag,graphicx}
\input epsf

\def\gappeq{\mathrel{\rlap {\raise.5ex\hbox{$>$}}
{\lower.5ex\hbox{$\sim$}}}}
\def\lappeq{\mathrel{\rlap{\raise.5ex\hbox{$<$}}
{\lower.5ex\hbox{$\sim$}}}}

\def\I{\rm 1\kern-.24em l}  

\begin{document}
\topmargin -1.0cm
\oddsidemargin -0.8cm
\evensidemargin -0.8cm

\pagestyle{empty}
\begin{flushright}
IFT-P.012/2005\\
July 2005
\end{flushright}
\vspace*{5mm}

\begin{center}
\vspace{3.0cm}
{\Large\bf
Relic Abundance of Mass-Varying Cold Dark Matter Particles }\\
\vspace{3.0cm}
{\large Rogerio Rosenfeld}\\
\vspace{.6cm}
{\it {Instituto de F\'{\i}sica Te\'orica - UNESP, Rua Pamplona, 145, 01405-900, 
S\~{a}o Paulo, SP, Brazil}}\\
\vspace{.4cm}
\end{center}

\vspace{1cm}
\begin{abstract}
In models of coupled dark energy and dark matter the mass of the dark matter
particle
depends on the cosmological evolution of the dark energy field. 
In this note we exemplify in a simple model the effects of this mass 
variation on the relic abundance of cold dark matter.
\end{abstract}

\vfill
\eject
\pagestyle{empty}
\setcounter{page}{1}
\setcounter{footnote}{0}
\pagestyle{plain}
%



We still do not know the origin and composition of the cold dark matter (CDM) 
in the universe. 
Recent precision measurements of the cosmological microwave background
and of the large scale structure of the universe 
put a strict bound on the abundance of non-baryonic CDM \cite{wmap}:
\begin{equation}
\Omega_{CDM} h^2 = 0.113 \pm 0.008, 
\end{equation}
resulting from a fit of several measurements combined in the framework of a
$\Lambda$CDM model with a running spectral index.

Most probably CDM is
made of particles (even though there are alternatives where dark matter is the
manifestation of a fluid with a non-standard equation of state, such as the
Chaplygin gas and quartessence models \cite{quartessence}) and the default 
candidate is the 
lightest supersymmetric particle
(LSP) of supersymmetric (SUSY) extensions of the standard model of electroweak
interactions, which is stable if R-parity is conserved.

The calculation of the LSP abundance in the universe has reached a very
sophisticated level. Computer codes are now publicly available that take into 
account all
the several LSP annihilation and co-annihilation processes that enter in the evaluation of its
abundance today \cite{codes}.
The comparison of the results of the computer codes with recent observations are
placing strong constraints in the parameters of SUSY 
(actually, minimal supergravity models) \cite{constraints}. In fact, accuracies
of the order of $10 \% $ among differents codes are being sought
\cite{comparison}.

We also know that the universe is accelerating today. There should exist a form of dark energy, comprising roughly $70 \%$ of the energy density of the universe, 
responsible for its acceleration. The simplest possibility, still consistent with
cosmological data, is a cosmological constant. One tantalizing problem that arises in these
models is the so-called coincidence problem: why dark energy starts to dominate the universe
only at recent times? This has motivated the study of models in which 
dark energy is coupled to dark matter \footnote{There are also recent models
that couples dark energy to neutrinos so that the energy density of neutrinos
tracks the dark energy density \cite{neutrinos}}. 

In these models of coupled dark energy, the mass of the dark matter particle depends on the
dark energy field and therefore it varies on a cosmological time scale. 
In this letter we point out in a general way what consequences this effect may have 
in the calculations of the cosmological abundance of cold dark matter, 
illustrating them with a particular simple model.

There are several different models of coupled dark energy, sometimes referred to as VAMPs (VAriable-Mass Particles) in the literature. The mass of
the dark matter particles evolves according to some function
of the dark energy field $\phi$, as, for example, a linear function of the field
\cite{carrolvamps, quirosvamps,  peeblesvamps, hoffmanvamps} with
a inverse power law dark energy potential or
an exponential function
\cite{exp1amendola, exp2amendola, exp3amendola,pietronivamps,riottovamps,us}
with an exponential dark energy potential. For instance, if the dark matter is a fermion one could have
an interaction like $g(\phi) m_0 \bar{\chi} \chi$, where the fermion mass
$m_\chi = g(\phi) m_0$ is a function 
of the dark energy field.
Since the dark energy field is dynamically evolving with time, one would have 
$m_\chi = m_\chi(\phi(a)) = m_\chi(a)$. 

In order to have a rough idea of the possible magnitude of the mass-varying effect, 
we will follow a more phenomenological approach and assume that the dark matter 
particle mass evolves with the cosmological scale factor $a$  as \cite{majerotto}:
\begin{equation}
m_\chi(a) = m_\chi^{(0)} e^{\int_0^{\ln a} \delta(\alpha') d \alpha' }
\end{equation}
where $m_\chi^{(0)}$ is the particle mass today and $\alpha = \ln a$.
This results in the following equation for the evolution of the CDM energy density $\rho_{CDM}$:
\begin{equation}
\dot{\rho}_{CDM} + 3 H \rho_{CDM} - \delta(a) H \rho_{CDM} = 0
\end{equation}
where $H = \dot{a}/a$ is the Hubble parameter.
Conservation of the total stress-energy tensor them implies that the 
dark energy density should obey
\begin{equation}
\dot{\rho}_{DE} + 3 H \rho_{DE} (1 + w_{DE}) + \delta(a) H \rho_{CDM} = 0
\end{equation}
where $w_{DE}$ is the equation of state of the dark energy fluid.

Many coupled dark energy models presents a so-called 
scaling solution where the dark matter and dark energy densities 
scale as simple functions of the scale factor. Following Majerotto et al. \cite{majerotto}
we impose:
\begin{equation}
\frac{\rho_{DE}}{\rho_{CDM}} \propto a^\xi
\end{equation}
This condition implies in the following expression for $\delta$:
\begin{equation}
\delta = \frac{\delta_0}{\Omega^{0}_{DE} + (1- \Omega^{0}_{DE}) a^{-\xi}}
\end{equation}
where
\begin{equation}
\delta_0 = -\Omega^{0}_{DE} (\xi + 3 w_{DE})
\end{equation}

The free parameters of this model are taken to be $\Omega^{0}_{DE}$, $w_{DE}$ and 
$\delta_0$. From a fit with the supernova Ia gold sample \cite{riess}, Majerotto et al.
obtain \cite{majerotto}:
\begin{equation}
\Omega^{0}_{DE} = 0.62; \;\;\;
w_{DE} = -1.9; \;\;\;
\delta_0 = -1.5.
\end{equation}
which implies in $\xi = 8.1$.

With these values we can illustrate the effects of cold dark matter mass 
variation and we find that 
it saturates rather quickly for $a < 0.1$ at the value:
\begin{equation}
\frac{m_\chi(a)}{m_\chi^{(0)}} = 1.34
\end{equation}

In order to estimate the consequences of the mass variation on the relic abundance of cold dark matter
we have to solve the Boltzmann equation for the comoving number density 
$Y = n/s$, where $n$ is the number density and $s$ is the 
entropy density, as a function of the variable $x = m(a)/T$:
\begin{equation}
\frac{x}{Y_{EQ}} \frac{d Y}{d x} = 
- \frac{\Gamma_A}{H} \left[ \left( \frac{Y}{Y_{EQ}} \right)^2 -1 \right]
\end{equation} 
where $Y_{EQ}$ is the equilibrium distribution and 
$\Gamma_A = n_{EQ} \langle \sigma_A v \rangle$ is the total annihilation width 
written in terms of 
the thermally averaged annihilation cross section times the relative velocity.
The approximate solution of the Boltzmann equation gives the 
following relation for the CDM abundance today \cite{kolbturner}:
\begin{equation}
\Omega_{CDM} \propto Y_\infty m_\chi^{(0)}
\end{equation}
where $ Y_\infty$ is the solution of the Boltzmann equation for large $x$, when
the number density is frozen after decoupling. For an $s$-wave annihilation
process, which we will consider below, $ \langle \sigma_A v \rangle = \sigma_0$
and the solution can be well approximated by
\begin{equation}
Y_\infty \propto \frac{x_F}{m_\chi^{(F)} \sigma_0}
\end{equation}
where $x_F$ is the freeze-out value of the variable $x$ and 
$ m_\chi^{(F)}$ is the CDM mass at freeze-out. We have kept only terms which have
dependence on the CDM mass. In fact, in the following we will neglect a
logarithmic mass dependence in $x_F$ to obtain
\begin{equation}
\Omega_{CDM} \propto  \frac{m_\chi^{(0)}}{m_\chi^{(F)}} \frac{1}{\sigma_0}
\end{equation}

At this point, the only particle physics assumption was that of CDM 
$s$-wave annihilation. Therefore, naivelly one would think 
that there is a simple linear difference due to the possibility 
of mass-variation.

However, the dependence can be potentially larger.
As an example, we will consider a  non-relativistic annihilation process 
$\bar{\chi} \chi \rightarrow \bar{f} f$ with a vector interaction with a coupling 
constant $g$  mediated by a vector boson of mass $M_V$. 
In this simple case, away from
the resonance (for $M_V \gg m$) we have
\begin{equation}
\sigma_0( \bar{\chi} \chi \rightarrow \bar{f} f) = 
\frac{1}{2 \sqrt{2} \pi} \frac{g^4}{M_V^4} (m_\chi^{(F)})^2 (1 + v^2 \beta_f^2/3)
\end{equation}
where $\beta_f = \sqrt{1- m_f^2/(m_\chi^{(F)})^2}$. The important fact is the dependence on
$(m_\chi^{(F)})^2$. Hence one would get:
\begin{equation}
\Omega_{CDM} = \left(\frac{m_\chi^{(0)}}{m_\chi^{(F)}} \right)^3
\Omega_{CDM}^{\mbox{\small c. m.}}
\end{equation}
where $\Omega_{CDM}^{\mbox{\small c. m.}}$ 
is the usual result of the CDM abundance today with a constant mass particle. 

Therefore, in this simple illustrative model a difference of a factor of
$(1/1.34)^3 = 0.41$ in the cold relic 
energy density can be easily obtained.    
In these times of precision measurements and ruling out of parameter space 
for well defined models, the varying-mass phenomenom can be of importance.
However, one should keep in mind that in realistic models there are several
annihilation channels with different dependence on the CDM particle mass
\cite{JKG}.

In summary, models that couple dark energy with dark matter predict mass-varying dark matter particles.
Here we have exemplified the effect of mass variation in a simple model. Of course any realistic model can 
be solved to find the function $m(a)$ and verify the effect on the abundance.
It would be interesting to implement the possibility of dark matter mass variation in 
realistic computational codes of relic abundances.

\section*{Acknowledgments}
I would like to thank Luca Amendola, Urbano Fran\c{c}a and Sasha Belyaev for 
useful comments.
This work is partially supported by a CNPq grant. 
%
%

\end{document}